**Towards Automating Structural Discovery in Scanning Transmission Electron Microscopy**

Nicole Creange,[1,a] Ondrej Dyck,[1] Rama Vasudevan,[1] Maxim Ziatdinov,[1,2] and Sergei V. Kalinin[1]

[1] Center for Nanophase Materials Sciences, Oak Ridge National Laboratory, Oak Ridge, TN 37831

[2] Computational Sciences and Engineering Division, Oak Ridge National Laboratory, Oak Ridge, TN 37831

**Abstract**

Scanning transmission electron microscopy (STEM) is now the primary tool for exploring functional materials on the atomic level. Often, features of interest are highly localized in specific regions in the material, such as ferroelectric domain walls, extended defects, or second phase inclusions. Selecting regions to image for structural and chemical discovery via atomically resolved imaging has traditionally proceeded via human operators making semi-informed judgements on sampling locations and parameters. Recent efforts at automation for structural and physical discovery have pointed towards the use of 'active learning' methods that utilize Bayesian optimization with surrogate models to quickly find relevant regions of interest. Yet despite the potential importance of this direction, there is a general lack of certainty in selecting relevant control algorithms and how to balance *a priori* knowledge of the material system with knowledge derived during experimentation. Here we address this gap by developing the automated experiment workflows with several combinations to both illustrate the effects of these choices and demonstrate the tradeoffs associated with each in terms of accuracy, robustness, and susceptibility to hyperparameters for structural discovery. We discuss possible methods to build descriptors using the raw image data and deep learning based semantic segmentation, as well as the implementation of variational autoencoder based representation. Furthermore, each workflow is applied to a range of feature sizes including NiO pillars within a La:SrMnO$_3$ matrix, ferroelectric domains in BiFeO$_3$, and topological defects in graphene. The code developed in this manuscript are open sourced and will be released at github.com/creangnc/AE_Workflows.

[a] creangenc@ornl.gov

# I. INTRODUCTION

Over the last several decades, electron microscopy has become the mainstay of multiple fields ranging from materials science to condensed matter physics and biology.[1-3] In particular, the introduction of aberration correction in late 90's and its broad commercialization in subsequent decades has enabled advancements ranging from high veracity structural imaging,[4-6] electron energy loss spectroscopy (EELS) with single atom sensitivity,[7] and depth-resolved imaging.[8,9] The last decade has significantly expanded this range of capabilities due to emergence of sub-10 mV resolution EELS of plasmonic and phonon excitations,[10] 4D scanning transmission electron microscopy (STEM) imaging,[3,11-13] and beams with orbital momentum.[14,15] Similarly, momentous progress has been made in dynamic STEM imaging, allowing visualization of thermally, chemically, and beam induced transformations.[16,17] Recently, these beam induced reactions were harnessed to enable patterning,[18] manipulation,[19,20] and assembly[21,22] at the atomic level. However, despite the gamut of advances in the development of source and detectors in STEM, the fundamental principle of scanning microscopies has remained largely the same, namely the sequential acquisition of data and the data collection workflow.

In structural STEM imaging, the beam is continuously rastered over a square region of interest. The detected signal can be plotted over the 2D grid, giving rise to conventional STEM images. In hyperspectral imaging modes, such as EELS imaging or 4D STEM, the beam is positioned over the centers of a rectangular grid, yielding the 3- or 4D data set over a uniformly sampled spatial grid. These imaging modes offer the advantage of the ease of implementation and convenient data visualization.

From the information theory viewpoint, the uniform sampling grid corresponds to the case of uniform Bayesian prior, or lack of any prior knowledge of the system.[23,24] This choice is fully justified in the case of the fully unknown sample. However, in most materials system the distribution of data of interest is non-uniform in the image plane. For example, in ferroelectric thin films, the phenomena of interest occur at interfaces and topological defects which appear in non-uniform patterns. In nanoparticle systems, phenomena of interest occur at the individual nanoparticles and especially at their junctions, whereas the empty space between them is often featureless. Finally, in atomically resolved data of systems such as graphene and 2D chalcogenides, phenomena of interest occur at atomic and extended defects, which can occupy a relatively small fraction of the image plane. Correspondingly, an optimal imaging strategy will preferentially discover these physically-relevant features.

Human-operated STEM workflows are the mainstay of the microscopy field; however, this process is extremely limiting. First and foremost, the latency time of a human operator is well below that of the image data acquisition in STEM. Hence, identification of regions of interest is currently human limited, making studies of dynamic phenomena such as catalysis, growth, or electron-beam manipulation slow. Secondly, the human eye is generally well suited for the identification of certain localized features, but the selection is highly biased by the operator experience and interest, introducing a large degree of observation bias in the exploration. Similarly, the discriminative capability of the human eye is limited and can be strongly affected by the chosen visualization settings such as image contrast and color scale. Finally, subtle changes in the image, such as a slight periodicity change, are usually undetected by the human eye and can be revealed only by post-experiment image analysis.

These considerations, as well as recent advances in autonomous robotics and vehicles, have sparked a tremendous interest towards the development of automated experimentation (AE) in general microscopy[25-27] and STEM[28,29] in particular. Some of the specific targets for AE include the identification and subsequent imaging of specific objects of interest in static systems, aiming at: minimizing the total measurement time and consequent beam damage, imaging dynamic systems, and ultimately controlled/completely automated probe manipulation. In the most general case, implementation of the AE in microscopy requires a set of synergistic instrumental and control advances, including controlled beam/probe motion, rapid analysis of the data to extract descriptors of interest, and feedback based on these descriptors.

For scanning probe microscopies, many of the instrumental aspects of these developments are far from trivial, however, control of the probe scan path was demonstrated fairly early by Schwartzentruber[30] for scanning tunneling microscopy (STM) and Ovchinnikov et al.[31] and Requicha et al.[32-34] for atomic force microscopy (AFM). Furthermore, broad implementation of non-rectangular scans has started over the last several years,[35-37] paving the way into the development of automated processes in AFM and STM. These developments in turn necessitate the development of mathematical tools that can convert the data acquired over complex scan patterns into the standard 2D images, including compressed sensing[35,36,38,39] and Gaussian Process[40] methods. Due to the computational complexity, this has become viable only over the last five years. Comparatively in STEM, the progress has been slower due to the closed hardware architecture of commercial microscopes.

The key remaining element is the identification of the structures of interest and generation of the feedback signal. It was realized for atomic fabrication based on simple FFT,[22] but this specific example does not generalize readily for more complex cases. Despite the relatively small number of demonstrations, currently the enabling instrumentation and software controls for automated experiment in STEM, if not necessarily widespread, are at least accessible. The proliferation of the software environments such as NION SWIFT[41] or PyJem[42] further suggest that soon automated experiments will become commonplace. Missing now are the algorithms and workflows that allow implementation of automated experiments in STEM, i.e. convert the video stream from the detector into the control signals to the beam scanning, optimizing (in a certain sense) information acquisition.

Here, we discuss several general paradigms for the automated experiment in STEM utilizing classical rectangular scan workflows and develop the experimental workflow for structural discovery. In this discussion, we consider the factors such as the existence of the partial or full prior knowledge on the system, knowledge of the exploration targets, the nature of the collected signal used to guide exploration, and the presence and relevance of the dynamic changes in the system. We further discuss how these approaches can generalize to the non-rectangular predefined scans and freeform scans.

## II. METHODS

The simplest form of AE in imaging is when the objects of interest are known *a priori*, from prior experiments or theoretical considerations. These can be formulated based both on the positive criteria (known atomic defect configurations, ferroelectric domain walls) or negative criteria (i.e. breaking of ideal lattice). In this case, the identification of the object of interest can be based on a simple function that converts the local signal vector into the measurement of interest. The signal vector can be either the sub-image, including classical rectangular scan, the scan over an arbitrary preset trajectory, or the scan over a dynamically adjusted trajectory, or a spectrum. For scans over a rectangular or static preset trajectory, the conversion from image data to the measurement of interest can be performed using the suitably chosen function, e.g. based on a simple correlation function or a more complex deep convolutional neural network.

In the cases where the features of interest are unknown, the exploration algorithm can be based on the dynamic libraries of objects of interest. In this approach, the microscope exploits the unknown system and identifies the anomalies using the initial batch of experimental data. These are used to create the features of interest in the unsupervised manner. With these, the physical or statistical features of interest can be used to guide the probe motion over the image plane, e.g. using Bayesian Optimization. Here, we develop the approach for automated experiment workflow in STEM based on a two-tiered approach including supervised atom identification via deep convolutional neural networks and unsupervised feature of interest discovery via rotationally invariant variational autoencoders. The autoencoder output is used as the surrogate model for the acquisition function which guides the Bayesian optimization algorithm. This approach is compared with the simpler method based on the pretrained linear transform-based AE workflows.

As a static data set, presented in Figure 1, we chose high-angle annular dark-field images of various materials which each contain different feature lengths. The dataset includes: NiO pillars in a La:SrMnO$_3$ (NiO-LSMO) matrix,[43-45] sample of BiFeO$_3$ (BFO),[46-48] Si-containing graphene.[49,50] Each of these images contain different information of interest, such as domain walls, defects, and dopants. In the NiO-LSMO system, the phases are obvious, however it is unknown if all the interfaces are the same or if any structural anomalies lie in the bulk. For Sm-BFO and graphene, the aim is to automatically identify topological defects but at different length scales.

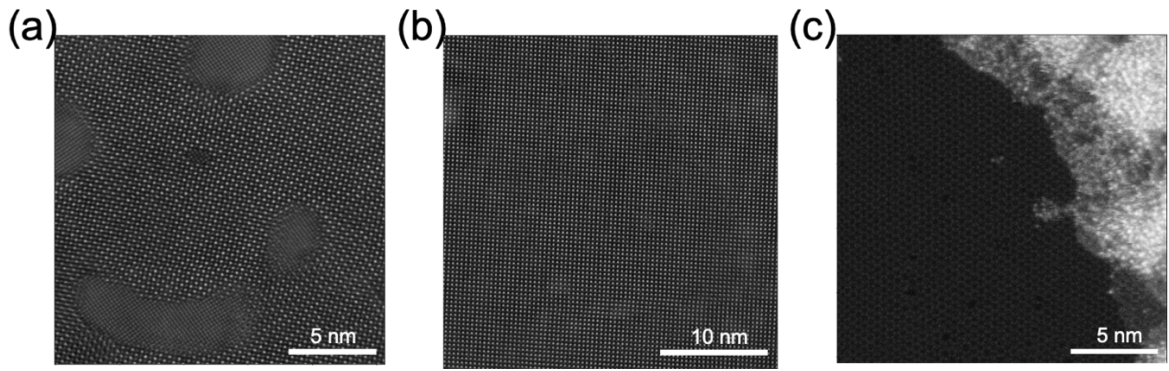

Figure 1: High-Angle annular dark-field (HAADF) STEM images of (a) NiO pillars in La:SrMnO$_3$,[43,44] (b) BiFeO$_3$[46-48] and (c) graphene.[49,50]

The proposed analysis workflows are generally separated into three parts: image segmentation, linear or non-linear dimensionality reduction, potentially allowing for physics-based invariances, and data collection. Within each part, a variety of methods exist which may produce similar results. Here we will focus on two methods of image segmentation and two methods of image deconvolution while the data collection is simulated via Bayesian optimization.

For image segmentation, the first method makes use of pre-trained deep convolutional neural networks (DCNN). This choice is dictated by the fact that in most cases of interest in the atomically resolved STEM and STM studies are the structures and transformations on the atomic level. Furthermore, DCNN based atom finding is robust and hence can be performed given the pretrained networks for a broad range of image sizes, sampling, signal to noise ratios, etc. The DCNN predictions can further be improved via ensemble learning and iterative training approaches. It is important to also note that DCNN can be trained for the semantic segmentation of mesoscopic images for determination of specific microstructural elements, e.g. particle and grain boundary finding, determination of the ferroelectric domain walls, etc., allowing for generalization of this approach towards other (prior known) objects of interest.

The output of the DCNN is the softmax probability density that a given image pixel belongs to atom (single layer) or specific atomic class. This softmax probability density, different from a Bayesian posterior density or ensemble uncertainty, can be further used to reconstruct the centroids of the atoms, and hence determine the atomic coordinates. The latter can be further refined, either from original or semantically segmented images, using the functional fits or more complex Bayesian inference methods that include full or partial knowledge of the beam shape. Based on the atomic coordinates, the local neighborhood of the selected atom[51] can be used as a local structural descriptor vector.

The alternative image segmentation approach can be based on a suitable transform of the sub-image at an arbitrary position (i.e. not necessarily centered on atom). This can be performed via fast Fourier Transform (FFT) of the sub-image[52], or Radon or Hough transform of the mesoscopic data. However, in this case selection of the surrogate model requires prior knowledge of the exploration target. For example, this can be achieved via projection of the FFT on a specific linear unmixing component known *a-priori*, or selection of the intensity at a given coordinate. In the proposed workflows, we will explore the use of a sliding window FFT to produce a stack of sub-images.

The second stage of the general workflow involves image deconvolution in which two approaches are discussed: manual decomposition via non-negative matrix factorization/single value decomposition and unsupervised rotationally invariant variational autoencoders (rVAE). Briefly, variational autoencoders refer to the general class of the encoder-decoder networks in which the encoder projects the large dimensional vector space to a small number of latent variables. The decoder then draws variables from the latent space and up-samples them to the image. The encoder and decoder are trained jointly, providing for the efficient encoding of input object as a continuous latent variable. The rotationally invariant VAEs are configured so that three of the latent variables are rotational angle and x- and y- offsets, allowing to localize similar objects at different orientations within the image plane.

Both approaches produce low-dimensional representations which are then mapped back onto the image and selection of the appropriate dimension which represents the property of interest can be selected. On transition to the AE stage, the selection of the latent variable(s) is utilized to define the surrogate model used in BO. In the simplest case, the latent variable with the largest change across the latent space can be used; however, here the human intervention can be used to select the latent variable, or a combination of auto-selection and human selection can be used. Note that while VAEs belong to the unsupervised machine learning methods, they rely on the existence of a full data set for training. This is not the case for AE, where the information becomes accessible through the course of the experiment. Here, we explore the approach where the rVAE is trained during the discovery stage of the experiment.

We further implement the Bayesian optimization methods for AE with Gaussian Process (GP) as the surrogate models. Generally, GP refers to an approach for reconstruction of relationships between two sets of parameters and yields the predicted values of a function and the associated uncertainty. This surrogate model is then used in conjunction with an acquisition function during the Bayesian optimization process. In this whole process, we have several factors which control the relationship between parameters including the GP covariance function (commonly referred to as kernel) and the type of acquisition function. The choice of kernel sets the base 'shape' of the relationship and common kernels include Gaussian and Matern kernels.[53]. The search strategy is based on the balance between exploration and exploitation. In the exploration regime, we are able to search a large area of the sample for regions of interest, whereas in the exploitation regime, we probe near an existing area of interest to gain more information around the area. Here we use three different acquisition functions: expected improvement, upper confidence bound, and probability of improvement. The feature length scale and particular data set are critical to the specific kernel and acquisition function used. For example, well defined features and large feature sizes could make use of a number of different kernels and acquisition functions and yield very similar results whereas a small feature which is highly local may see vastly different results with a change in the kernel or acquisition function.

Our proposed workflows, outlined in Figure 2, are comprised of three different pathways for the automation of data, with varying degrees of human supervision: 1) sliding window Fast Fourier transform (FFT), dimensionality reduction via single value decomposition (SVD), non-negative matrix factorization (NMF) or N-FINDR[54], component selection, BO (green line), 2) sliding window, rVAE, latent variable selection, BO, (green to orange line) and 3) DCNN, rVAE, latent variable selection, BO (purple line). Note that rVAE analysis can be performed with and without atom finding. These workflows vary in effectiveness depending on the feature length and are performed on three datasets with various feature lengths.

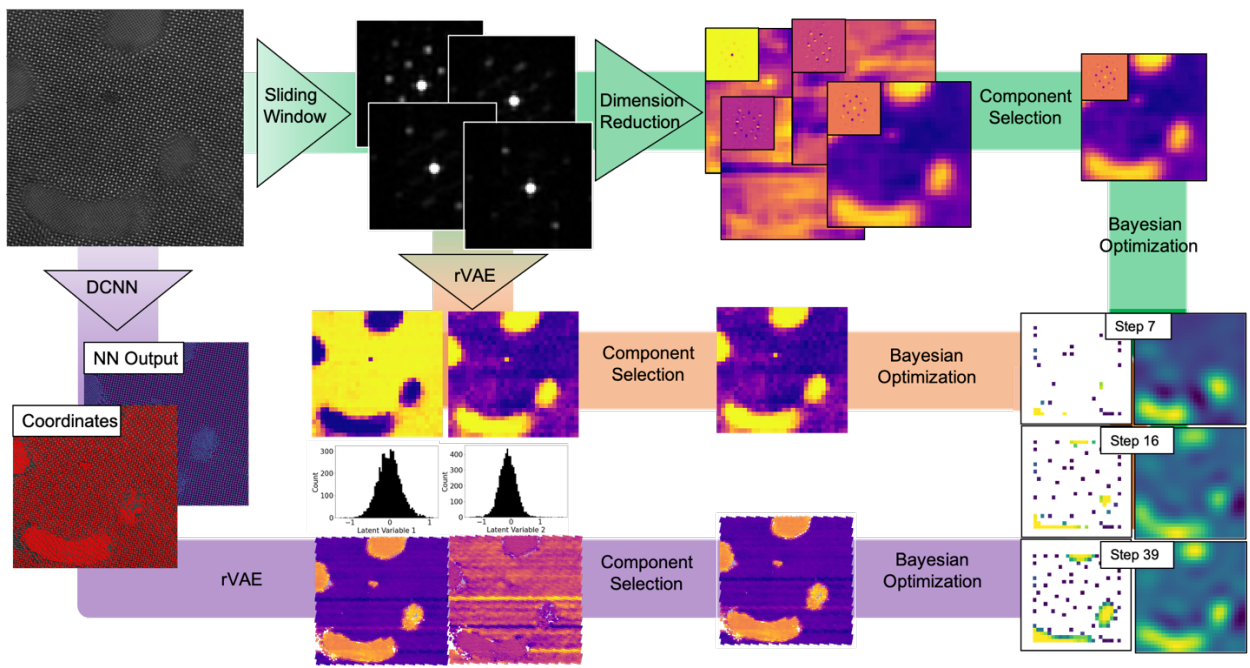

Figure 2: Workflow for the training and discovery phase for automated experiments with three different methods: human driven (green), partial human drive (orange), and computer driven (purple).

## III. DISCUSSION

The selection of a workflow will be highly dependent on the individual dataset; however, some trends of each workflow have been identified. Throughout all three datasets, workflow 1 and 3 consistently identified features of interest. The use of rVAE was found to be less reliable as the existence of multiple minima hindered convergence of the network.

Figure 3 highlights the component selection process for NiO-LSMO, Sm-BiFeO$_3$, and graphene using a primarily human driven workflow, workflow 1. The selection of a component which is representative of the feature of interest is key as this component will inform the BO process. In workflow 1, the image decomposition into individual components is performed using N-FINDR[54] for NiO-LSMO and Sm-BiFeO3, whereas NMF is used for graphene. It was noted that N-FINDR becomes less useful for the selection of small features, in which case a simpler image decomposition such as SVD or NMF can be used. Although the use of Bayesian unmixing, SVD, and NMF are simple decomposition techniques, they appear to be robust for feature isolation; however, human decision is necessary for the selection of technique, window size, and eventual component selection.

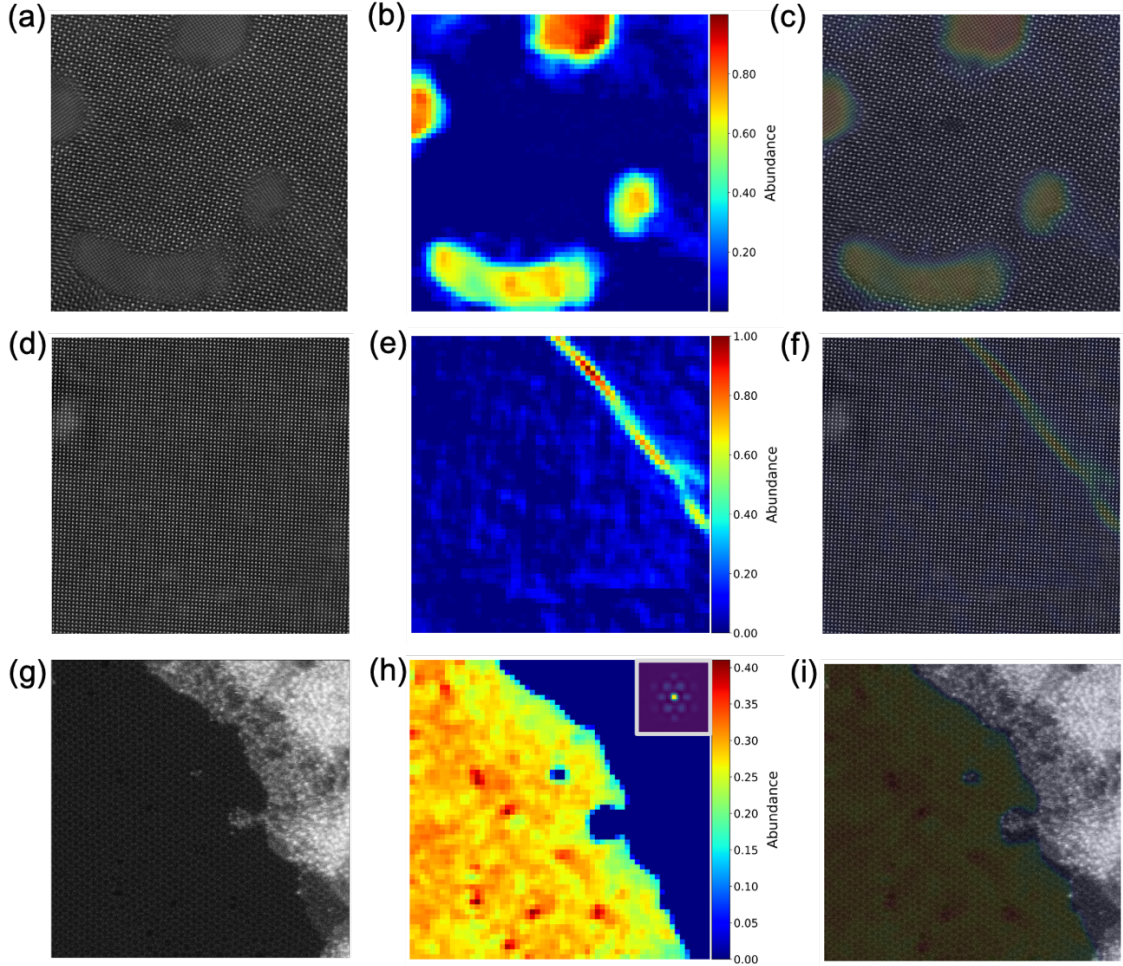

Figure 3: Transition from raw images of (a) NiO-LSMO, (d) BiFeO$_3$, and (g) graphene to abundance maps of an extracted component via (b,e) N-FINDR or (h) NMF and (c,f,i) their corresponding overlays, which are used for feature discovery (workflow 1).

A similar process is performed using a sliding window FFT and rVAE, workflow 2, but the component extraction is less robust than the simple image decomposition techniques used in workflow 1. The lack of robustness stems from the lack of network convergence, as the latent space contains numerous minima and prevents full convergence. In datasets where rVAE convergence was obtained, the latent variables were not always directly interpretable. Figure 4 outlines the variable outputs from trained rVAE networks where in each case, the input data was obtained via a sliding window FFT. In simple datasets, such as the NiO-LSMO seen in Figure 4a, the latent variable clearly separates the NiO pillars from the LSMO matrix and thus the use of the latent variable seen in Figure 4b would provide easy identification of the different phases. On the other hand, the latent variable for graphene, seen in Figure 4e, is less interpretable. Overlays of the two datasets, Figure 4c,f, clearly show the distinction between the interpretability of the latent spaces. As such, the use of this particular workflow would best be suited for data with large feature sizes or well-separated features.

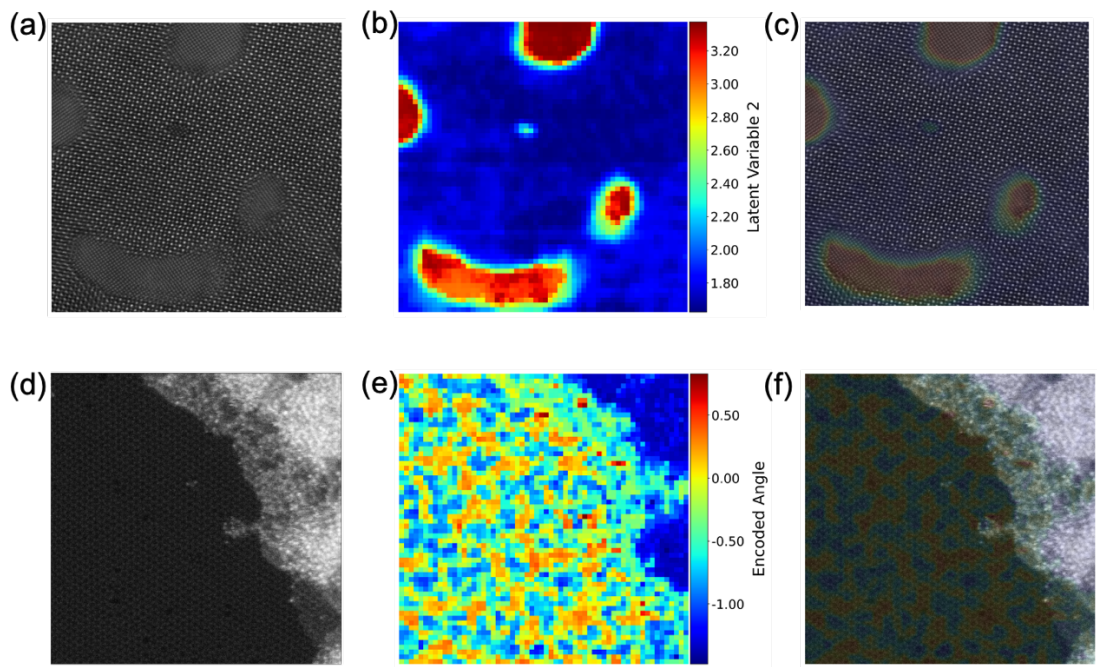

Figure 4: Transition from raw images of (a) NiO-LSMO and (b) graphene to (b,e) abundance maps of the latent variables found after rVAE training and (c,f) their corresponding overlays, which are used for feature discovery (workflow 2).

To obtain atomic level precision and realize semi-supervised automation, we utilized a DCNN for atom finding prior to use of the rVAE. As previously discussed, a pre-trained DCNN is implemented on a collected image in order to obtain each atomic column position. A sliding window is used to create a stack of images, each centered on an atomic column, which is then used for training of the rVAE network. Figure 5 reveals the latent variables found for each of the datasets used throughout this work. Here, the latent variables highlight clearly the separated phases of NiO-LSMO and the ferroelectric domain wall of Sm-BiFeO$_3$. The latent variable for graphene is less obvious though the encoded angle derived from rVAE is able to separate the two trimer orientations in graphene. The use of the DCNN and rVAE for feature identification is shown to be useful for both large and small features. Though, the use of a much simpler workflow, such as that of the sliding FFT window and NMF decomposition, may be of a more practical use for datasets with large feature sizes.
10

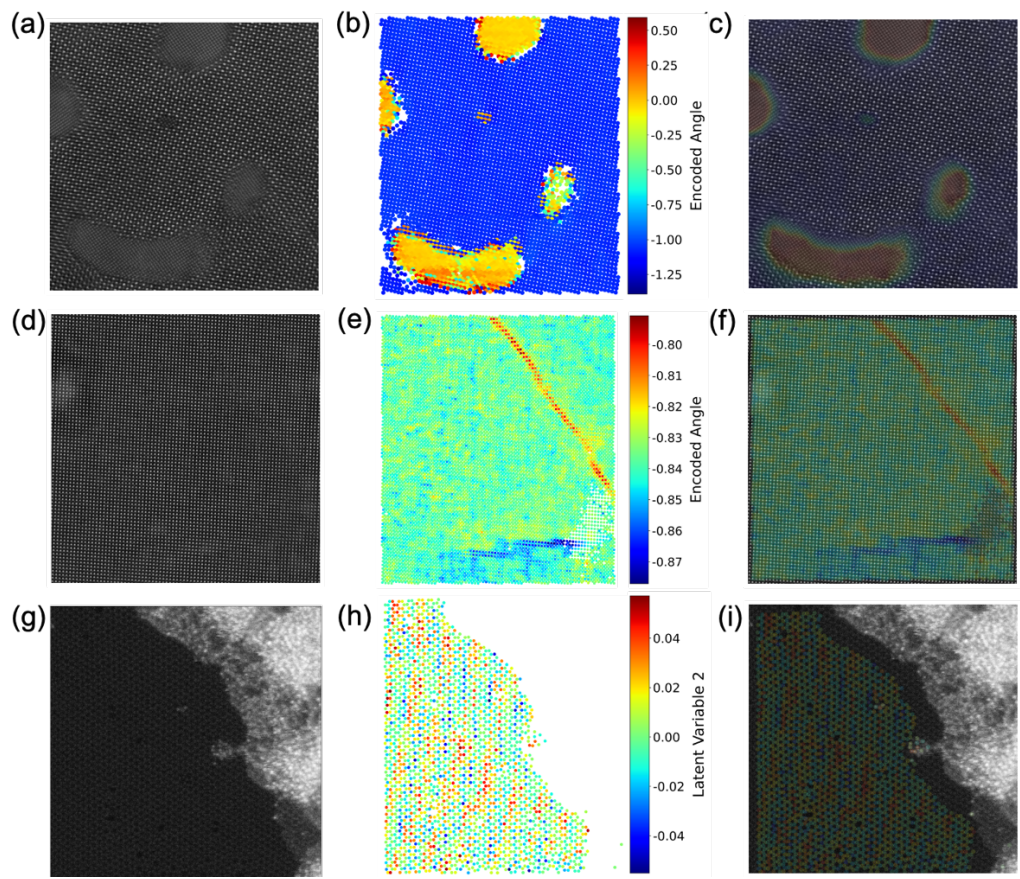

Figure 5: Transition from raw images of (a) NiO-LSMO, (b) BiFeO$_3$, (c) graphene to (b,e,h) abundance maps of the latent variables found after rVAE training and (c,f,i) their corresponding overlays, which are used for feature discovery (workflow 3).

After obtaining and selection a component or latent variable, it is used as the target for BO. Data points are collected according to the parameters within the BO algorithm and the image reconstruction and uncertainty are calculated using the data collected thus far. Figure 6 details the BO process for the NiO-LSMO system using the unsupervised workflow of a pre-trained DCNN and latent variable selection via rVAE. As these workflows are performed outside of a microscope, the latent variable map is used as the data feedback and reduced to a sparse grid of one data point to start the BO algorithm. Additionally, to speed up the data collection, batches of 10 points are used for each step of the BO process, such that at the 10$^{th}$ step, 101 data points have been collected. Upon implementing BO and the collection of 200-300 points, we are on average able to reconstruct the full original image with a low degree of uncertainty. Furthermore, most points collected are performed in the regions of interest. This decrease in collection time and collection of just around the feature of interest decreases the total beam time the sample is exposed to, ultimately decreasing the total beam damage. The precise number of points is highly dependent on multiple parameters including the feature size, the acquisition function, and the kernel.

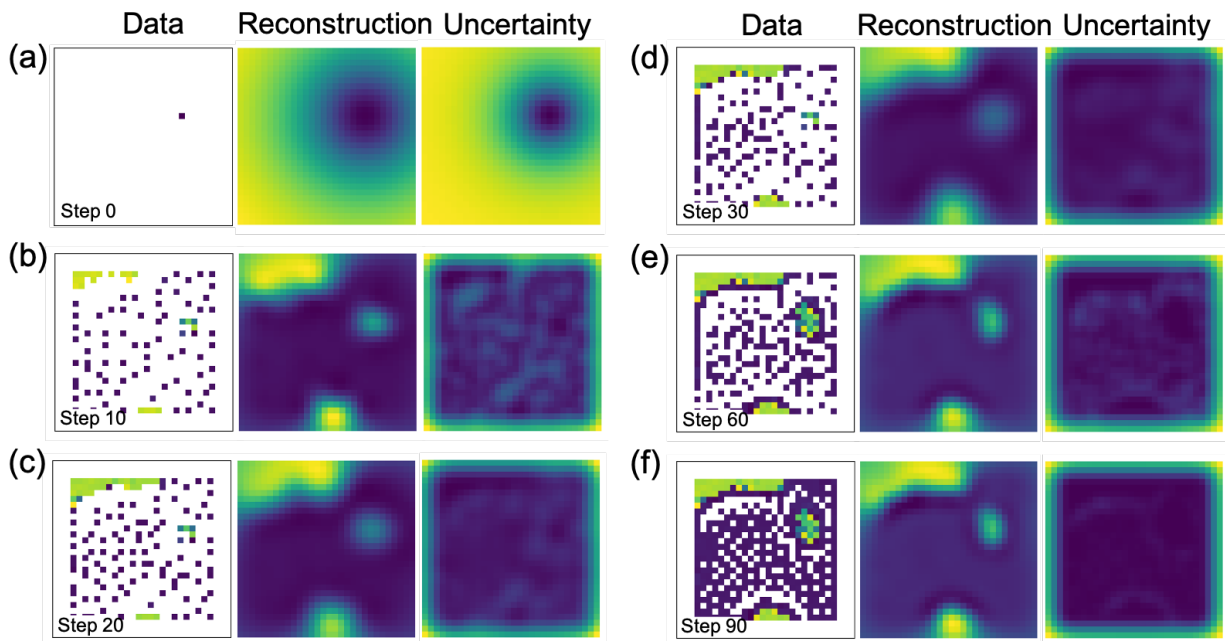

Figure 6: BO discovery for NiO-LSMO using Workflow 3. Shown are collection points, reconstructed images, and uncertainties after step (a) 0, (b) 10, (c) 20, (d) 30, (e) 60, and (f) 90 of the BO process.

We note that the acquisition function can have multiple close to degenerate maxima, particularly in the cases when large regions of image are uniform. While almost equivalent from the BO perspective, these maxima can differ strongly from the microscope perspective, i.e. certain measurements sequence can be preferential. For example, measurements of the near regions are easiest since probe motion is minimized, whereas scanning between opposite regions of image field can be sub-optimal. That said, it is obviously sub-optimal to scan very close to already scanned regions. Practically, this means that we need microscope-specific pathfinder functions that will allow balancing BO-derived points and microscope optimal performance.

These considerations suggest that a strategy is needed to configure the space. While multiple strategies can be explored, here we suggest a change in the correlation length (d), the kernel type, and the acquisition function parameters to strike a balance between exploration and exploitation. For each dataset considered here, combinations of correlation length, kernel type, and acquisition function were explored. We note that the correlation length will be highly dependent on the size of the data and the feature of interest size. In this work, the correlation length was considered between 1 and 20, the kernels consisted of RBF, Matern52, and Rational Quadratic, and the acquisition functions were varied between confidence bound, expected improvement, and probability of improvement. To evaluate the ability of full image reconstruction using the sparse dataset of features of interest, a structural similarity index measure (SSIM) is calculated at each parameter combination between the target data and the reconstructed image. Figure 7 outlines the SSIM and average GP uncertainty at the point of diminishing returns, where the collection of more data points does not improve the SSIM or the image reconstruction. While the full range of combinations have not been explored, those explored here provide insight into combinations which might work well for similar data sets.

From the BO parameter combinations explored thus far, we can observe that there is a wide spread of results for each dataset, though workflow 1 appears to be robust no matter the parameters used or the type of data. This robustness is lacking in the other two workflows seemingly due to the use of rVAE and the subsequent lack of network convergence. As workflow 1 does not require any network training, the performance in finding features of interest is steady throughout different types of data. However, the limitation appears in the level of quantification, as the sliding window FFT lacks the ability to obtain atomic level precision. When a dataset contains a feature of interest which is on the scale of single atoms, such as graphene defects or ferroelectric domain walls, workflow 3 performs adequately. Although the SSIM for workflow 3 is primarily below 0.5, the collection of data points is largely within the features of interest. The decrease is SSIM appears within the image reconstruction of small features, leading to a reconstruction which does not match the structural pattern of the target data.

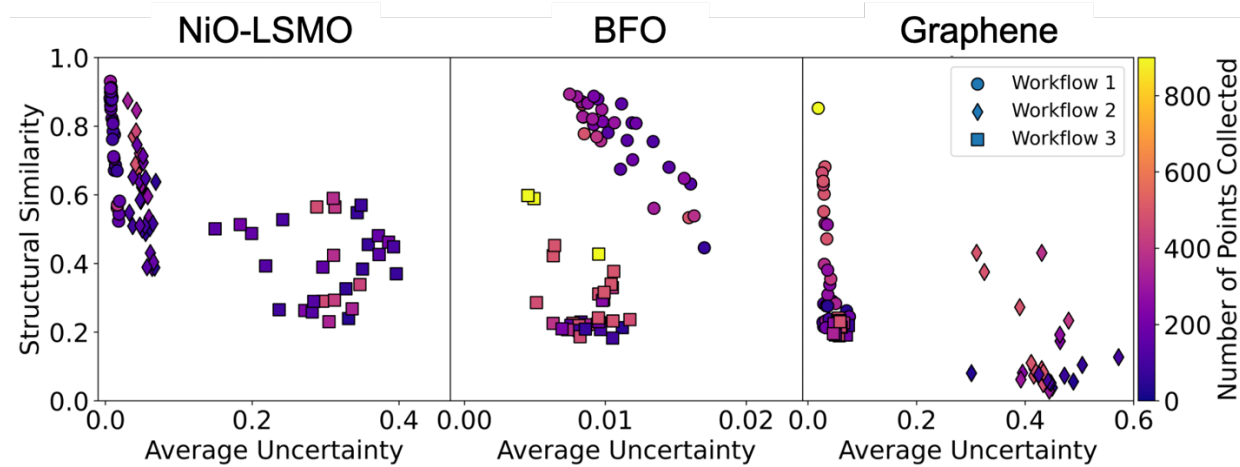

Figure 7: Structural similarity index as compared to the average uncertainty for the different workflows for the considered datasets.

The efficiency of each workflow is evaluated by finding the percentage of the feature of interest which is discovered at each BO step. The step at which the percentage is maximized is outlined in Figure 8, with each data point in a workflow and dataset represents a different set of hyperparameters. It is shown that workflow 1 has a robust nature in which, in most cases, over half of the feature of interest was discovered regardless of the hyperparameters used or the feature of interest size. Workflows 2 and 3 are highly variable in their efficiency, as the hyperparameters and the convergence of rVAE training can highly effect the outcome. However, given a large feature size or a well defined feature, such as the NiO pillars in LSMO, workflow 2 can be highly efficient, with 99% of the feature of interest being discovered in fewer than 200 points. Further exploration and calibration of the hyperparameters would produce similar results to those seen in NiO-LSMO data, with highly efficient data collection and minimal user oversight.

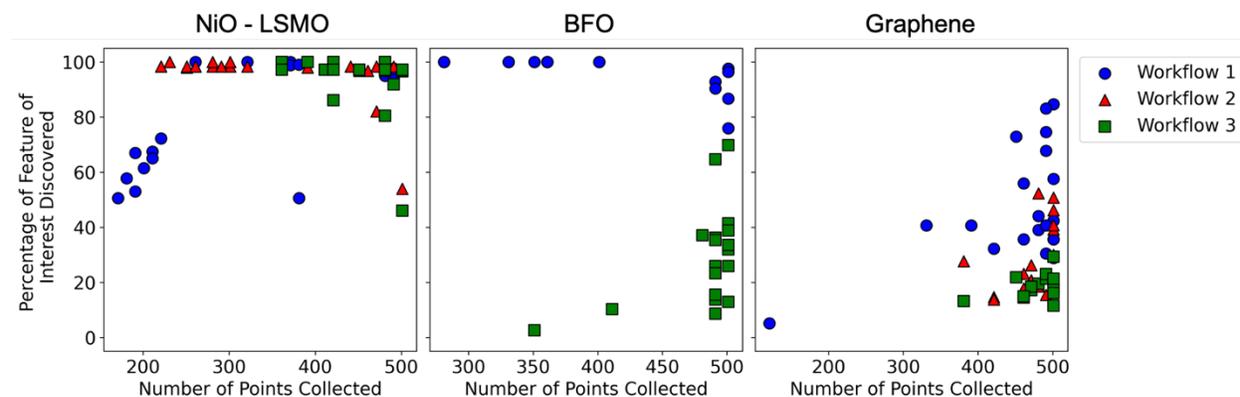

Figure 8: Reconstruction accuracy of BO algorithm presented as the percentage of feature of interest discovered and the associated number of points collected for each hyperparameter set for NiO-LSMO, BFO, and Si-graphene datasets.

Besides calibration of the workflows, the implementation on the microscope may present some complications. First, the use of a pre-trained DCNN may not work if the collected image is not within the distribution of data used to train the network. If this occurs, the network will need to be re-trained, or the data would need to be transformed to lie within the training distribution. Secondly, the collection of data within the BO process must be converted in a similar manner as the latent variable used. This could be a simple FFT transform or a more complicated process to obtain the latent variable obtained via rVAE. However, the manipulation of data can be trivial if the software environment which controls microscope is open source such as NION SWIFT[41] and PyJem.[42] Furthermore, given the software environment is accessible, the manipulation of the electron beam to specific locations determined via BO is trivial however there is some limitation in the speed of data collection when solely using the software environment. Nonetheless, the use of a field-programmable gate array (FPGA) connected directly to the microscope can bypass the software to enable the collection of data at a much faster rate and opens the avenue for real-time *in-situ* processing.

## IV. CONCLUSION

To summarize, broad incorporation of automated experiment in STEM and SPM necessitates development of machine learning algorithms that will guide the selection of the objects of interest. Given the specificity of STEM imaging, these approaches should take into account the generic features of the system, such as the presence of atomic columns in atomically resolved images or specific features in mesoscale images, and account for the presence of characteristic structural elements and patterns, such as phase separation and ferroelectric domains.

Here, we explored the workflows for AE in STEM which combined DCNN semantic segmentation and rVAE encoding in addition to a lower-level segmentation using sliding window FFT. On the training stage, the network defines and selects a specific parameter to explore, defined as the latent variable or component, while on the automated experimental stage, Bayesian optimization rapidly explores the sample space for features of interest. We note that the workflows can be made much faster, or we can obviate the training stage entirely via transfer learning, when the initial scans are used both for guiding exploration strategy and training rVAE.

Overall, the used of a specific workflow is dependent of the dataset and feature of interest. Workflow 1, consisting of a sliding window and image deconvolution, is robust and ideal for well-defined features of interest, while workflow 3, consisting of DCNN and rVAE, is best used for atomic scale features. For real deployment, there is a need to also introduce pathfinding functions, which balance the BO gain and the microscope scanning. These workflows can be generalized for different systems depending on the training of the DCNN. In addition, parameter tuning and main network training would need to be performed a-priori to data collection, as they are time-consuming; however, fine-tuning of the DCNN network could be performed within the workflow. Nevertheless, the workflows developed provide a gateway for rapid feature discovery, limited beam damage, and the collection of data primarily isolated within features of interest.

**Acknowledgements:**

This effort (ML and STEM) is based upon work supported by the U.S. Department of Energy (DOE), Office of Science, Basic Energy Sciences (BES), Materials Sciences and Engineering Division (O.D., S.V.K.) and was performed and partially supported (R.V.K., M.Z.) at the Oak Ridge National Laboratory's Center for Nanophase Materials Sciences (CNMS), a U.S. Department of Energy, Office of Science User Facility. We thank Jacob Swett for preparing and providing the graphene sample, Wenrui Zhang and Gyula Eres for the preparation and providing the NiO-LSMO sample, Matthew Chisholm for acquiring the NiO-LSMO STEM data, Ziaohang Zhang and Ichiro Takeuchi for preparing and providing the Sm-BFO sample, and Chris Nelson for acquiring the Sm-BFO STEM data.